\documentclass{iopjournal}
\usepackage{booktabs}
\usepackage{amsmath}
\usepackage{amssymb}
\usepackage{comment}
\begin{document}

\title{Simulation Design for Velocity-Controlled Spatio-Temporal Drivers in Laser Wakefield Acceleration} 

\author{Chiara Badiali$^1$\orcid{0000-0001-6450-7511}, Rafael Almeida$^1$\orcid{0000-0002-8843-5855}, Thales Silva$^1$\orcid{https://orcid.org/0000-0002-3061-6792} and Jorge Vieira $^{1}$\orcid{0000-0000-0000-0000}}

\affil{$^1$ GoLP/Instituto de Plasmas e Fus\~ao Nuclear, Instituto Superior T\'ecnico, Universidade de Lisboa, 1049-001 Lisbon, Portugal}

\email{chiara.badiali@tecnico.ulisboa.pt}


\begin{abstract}
Velocity-controlled spatio-temporal (ST) laser drivers offer a route to tailoring laser--plasma interactions by allowing the velocity of the intensity peak to be controlled independently of the envelope group velocity. In this work, we present a simulation-design workflow for PIC modelling of subluminal velocity-controlled ST pulses in OSIRIS based on a Maxwell-consistent spectral construction expressed as a superposition of exact vacuum solutions, and we describe its discrete $k$-space representation for numerical initialisation. We then examine wakefield excitation with velocity-controlled drivers, showing how the ST geometry couples the effective longitudinal extent of the high-intensity region to the transverse scale and deriving scaling guidelines for near-resonant excitation in the subluminal regime. Finally, we discuss the geometric constraints that make long-distance simulations costly, including focus--envelope slippage and strong transverse expansion, and we show that continuous wall injection can reproduce the intended vacuum propagation while substantially reducing the transverse domain size. Together, these results provide practical guidelines for accurate and computationally efficient PIC simulations of velocity-controlled ST drivers in wakefield-relevant regimes.
\end{abstract}
\section{Introduction}
\label{sec:intro}
Laser–plasma accelerators have emerged as a promising route towards compact sources of relativistic electron and ion beams~\cite{tajima1985high,RevModPhys.81.1229,PhysRevE.110.035202,adli2018acceleration} and secondary radiation~\cite{daido2012review,RevModPhys.85.1,galletti2024prospects,PhysRevLett.94.025003}, but their performance is often limited by dephasing and pump depletion~\cite{tajima2020wakefield,PhysRevX.9.031044}. A central challenge is therefore to control how the laser intensity is distributed in space and time as it propagates through plasma. Spatio-temporal (ST) pulse shaping, and in particular the \emph{ideal flying focus} concept, addresses this challenge by engineering laser pulses whose intensity peak travels at a prescribed velocity that is decoupled from the group velocity of the pulse envelope~\cite{sainte2017controlling,froula2018spatiotemporal,froula2019flying,yessenov2022space}. This control is achieved through chromatic focusing combined with a tailored spectral phase, allowing the focal spot to propagate with a chosen focal velocity $v_f$ (including subluminal and superluminal values) while maintaining an approximately constant peak intensity over distances that can greatly exceed the Rayleigh range of a conventional focus.

This added control has triggered strong interest in ST pulses across nonlinear optics and plasma physics~\cite{howard2019photon,franke2021optical}. Flying-focus and related ST structures have been proposed and, in some cases, demonstrated for driving ionisation waves with prescribed velocities~\cite{Caizergues2020NatPhot}, forming extended plasma channels~\cite{palastro2018ionization,turnbull2018ionization}, enabling dephasing-free laser wakefield acceleration, and tailoring wakefield structures for high-efficiency electron acceleration~\cite{PhysRevLett.124.134802,ambat2023programmable,miller2023dephasingless,ghasemi2024flying}. More recently, variants such as discrete and transverse flying-focus schemes have been introduced to enhance energy gain per stage or to accelerate ions, and structured beams have been used to directly drive wakefields in plasmas~\cite{Gong2024PRL,pierce2025laser,liberman2025electron}. In addition, subluminal ST drivers have been proposed as a means to phase-lock plasma wakefields to initially subrelativistic, short-lived particles and thereby extend the usable acceleration length in tailored-density plasmas~\cite{badiali2025plasma}. Overall, ST pulses with a prescribed focal velocity provide a flexible way to match the driver to the plasma response, rather than adapting the plasma to the constraints of a conventional laser pulse.

We focus on the subluminal regime because it is directly motivated by our studies of phase-locking wakefields to initially subrelativistic, short-lived particles using $v_f<c$ drivers~\cite{badiali2025plasma}. Nevertheless, several of the considerations discussed here, including the spectral construction and practical simulation-design guidelines, are also relevant to other operating regimes, such as $v_f\simeq c$ and $v_f > c$ schemes aimed at mitigating dephasing in laser--wakefield acceleration.

In this work, we address the physical and numerical requirements for modelling subluminal flying-focus-type ST pulses in particle-in-cell simulations, with emphasis on practical use within OSIRIS~\cite{2002Fonseca}. Building on the Maxwell-consistent spectral decomposition of~\cite{Almeida2023NonParaxialPIC}, we organise the paper around three main objectives: (i) presenting the spectral construction and its discrete $k$-space representation used for numerical initialisation; (ii) quantifying how the ST pulse parameters determine the conditions for resonant wakefield excitation; and (iii) characterising the geometric and numerical constraints that make these drivers computationally demanding, including focus–envelope slippage, transverse expansion, and the replica fields introduced by spectral discretisation.

The paper is organised as follows. Section~\ref{sec:theory_and_numerics} introduces the Maxwell-consistent spectral framework used to model velocity-controlled ST drivers and its numerical realisation in OSIRIS. Section~\ref{subsec:st_pulses} defines flying-focus pulses in Fourier space and distinguishes the subluminal and superluminal regimes through the geometry of the light-cone intersection. Section~\ref{subsec:numerical_init} describes how the continuous spectral weight is mapped onto a discrete set of modes for field initialisation in PIC codes. Section~\ref{subsec:periodicity} analyses the artefacts introduced by discrete $k$-space sampling and derives practical criteria to mitigate them.

Section~\ref{sec:design_validation} focuses on wakefield-relevant parameter choices and on the computational constraints that arise for subluminal drivers. Section~\ref{subsec:pulseshaping} discusses how the spatio-temporal geometry constrains the effective driver length and transverse scale, and provides scaling guidelines for near-resonant wakefield excitation. Section~\ref{subsec:wallinjectionadvantage} discusses the domain-size requirements imposed by envelope--focus slippage and the large angular content of velocity-controlled ST pulses, derives practical box-sizing estimates, and shows how wall injection enables long-distance simulations in reduced transverse domains. Finally, section~\ref{sec:conclusions} summarises the main conclusions and practical recommendations.

\section{Spectral construction and numerical implementation of spatio–temporal pulses}
\label{sec:theory_and_numerics}
Particle-in-cell codes commonly rely on paraxial beam models and the slowly varying envelope approximation to initialise laser pulses. While these approaches are adequate for narrowband Gaussian drivers, they are not suitable for velocity-controlled space–time (ST) pulses. The prescribed focal trajectory of ST pulses requires broadband spectral content, large-angle propagation components, and a fixed correlation between spatial and temporal frequencies. To address this limitation, we adopt a generalised, Maxwell-consistent initialisation in which the electromagnetic field is expressed as a superposition of vacuum solutions of Maxwell’s equations, thereby enabling accurate modelling of ST pulses. This section presents the mathematical formulation of this approach (section~\ref{subsec:st_pulses}), summarises its implementation in OSIRIS (section~\ref{subsec:numerical_init}), and discusses one of the issues arising from this method together with a strategy to circumvent it, namely the artificial periodic repetition of the fields induced by representing a continuous spectrum on a discrete $k$-space grid (section~\ref{subsec:periodicity}).

\subsection{Spectral design of spatio-temporal pulses}
\label{subsec:st_pulses}
In vacuum, any electromagnetic wave packet can be represented as a superposition of basis solutions that satisfy the vacuum dispersion relation
\begin{equation}
\omega(\mathbf{k}) = c|\mathbf{k}|.
\label{eq:light_cone}
\end{equation}
where $c$ is the speed of light in vacuum, $\omega$ is the angular frequency of the wave, and $\mathbf{k}=(k_x,k_y,k_z)$ is the wavevector. In Cartesian geometry, the electric field can be written as~\cite{Almeida2023NonParaxialPIC}
\begin{equation}
\mathbf{E}(\mathbf{x},t)=\Re\left\{E_0\int_{\mathbb{R}^3} f_k(\mathbf{k})\,
\hat{\mathbf{e}}_E(\mathbf{k})\,
\exp\!\left[i\left(\mathbf{k}\!\cdot\!\mathbf{x}-\omega(\mathbf{k}) t\right)\right]\,d^3\mathbf{k}\right\},
\label{eq:plane_wave_superposition}
\end{equation}
where $z$ defined as the main propagation direction, and $\hat{\mathbf{e}}_E(\mathbf{k})$ is a unit polarisation vector satisfying $\mathbf{k}\cdot\hat{\mathbf{e}}_E=0$. A corresponding expression can be written for the magnetic field. Equation~\eqref{eq:plane_wave_superposition}, together with the analogous expression for the magnetic field, guarantees that each spectral component is an exact solution of Maxwell’s equations in vacuum~\cite{sheppard1999electromagnetic,heyman2001gaussian}.  
The complex spectral weight $f_k(\mathbf{k})$ determines the amplitude and phase of each mode and therefore fully specifies the pulse, while the overall maximum amplitude of the electric field is set by $E_0$.

\begin{figure}[t]
 \centering
        \includegraphics[width=1.\textwidth]{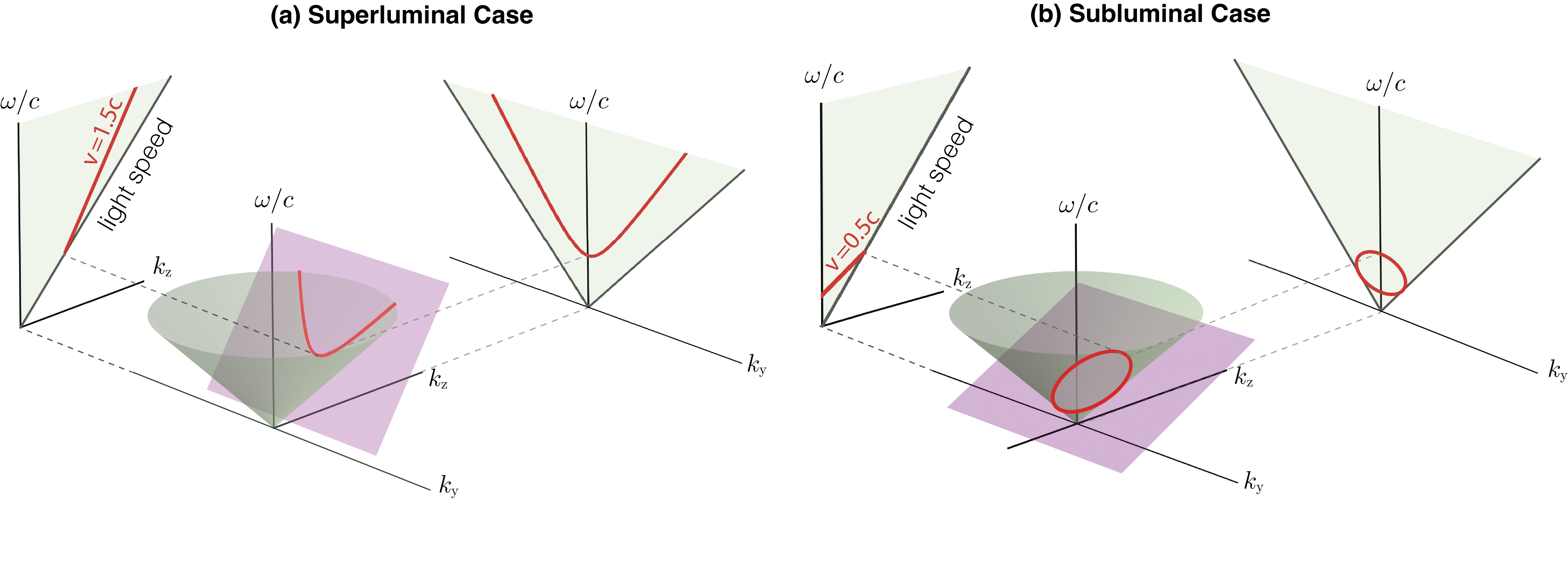}
    \caption{Spectral design of ST pulses with controllable focal velocities. The desired spectral components (red curve) are defined by the intersection of the vacuum light cone (green surface) with the plane imposed by the spatio-temporal coupling (purple plane).
    {(a)} In the superluminal regime ($v_f = 1.5c$), the intersection yields an open hyperbolic trajectory, characteristic of X-waves with infinite spectral extent.
    {(b)} In the subluminal regime ($v_f = 0.5c$), the coupling plane cuts through the time-like region of the cone, confining the allowed spectral range to a closed elliptical loop.}
\label{fig:spectral_stpulses}
\end{figure}
We now use equation~\eqref{eq:plane_wave_superposition} to model ST pulses. 
Figure~\ref{fig:spectral_stpulses} illustrates how a focal spot moving at a prescribed velocity $v_f$ can be generated. The allowed spectral components must be restricted to the intersection of the vacuum light cone, equation~\eqref{eq:light_cone}, shown in green in figure~\ref{fig:spectral_stpulses}, with a spectral plane defined by the desired trajectory,
\begin{equation}
\omega = v_f k_z + \left|\mathbf{k}_0\right| c \, (1 - v_f/c),
\label{eq:restrict_vf}
\end{equation}
shown in purple in figure~\ref{fig:spectral_stpulses}. The resulting intersection is highlighted by the red curve in figure~\ref{fig:spectral_stpulses}. Here, $(\omega_0,\mathbf{k}_0)$ defines the specific spectral plane under consideration, which passes through the point $\omega(\mathbf{k})=\omega_0$ when $\mathbf{k}=(0,0,k_0)$. In the $(k_z,\omega/c)$ plane, this curve is open in the superluminal regime ($v_f>c$), as shown in figure~\ref{fig:spectral_stpulses}(a), and becomes a closed loop in the subluminal regime ($v_f<c$), as shown in figure~\ref{fig:spectral_stpulses}(b). In addition, we restrict the spectrum to components with $k_z>0$ in order to describe forward-propagating pulses.

The intersection of equations~\eqref{eq:light_cone} and~\eqref{eq:restrict_vf} defines the curve $\kappa = 0$, where
\begin{equation}
\kappa = |\mathbf{k}| - k_0 - (k_z - k_0).
\end{equation}
To construct a physical pulse with finite energy, we allow the spectrum to have a finite bandwidth around $\kappa = 0$. For simplicity, we further assume a separable form for the spectral weight,
\begin{equation}
f_k(\mathbf{k}) = f_\parallel(\kappa)\, f_\perp(\mathbf{k}_\perp),
\label{eq:fk_def}
\end{equation}
where $\mathbf{k}_\perp=(k_x,k_y)$ denotes the transverse wavevector. Among the possible choices for $f_\perp$, we adopt a Gaussian transverse spectrum,
\begin{equation}
f_\perp(\mathbf{k}_\perp) = \frac{w_0^2}{4\pi}\exp\!\left(-\frac{k_\perp^2 w_0^2}{4}\right),
\label{eq:fk_perp}
\end{equation}
which corresponds to a beam waist $w_0$ in real space.

\begin{figure}[t]
 \centering
        \includegraphics[width=0.7\textwidth]{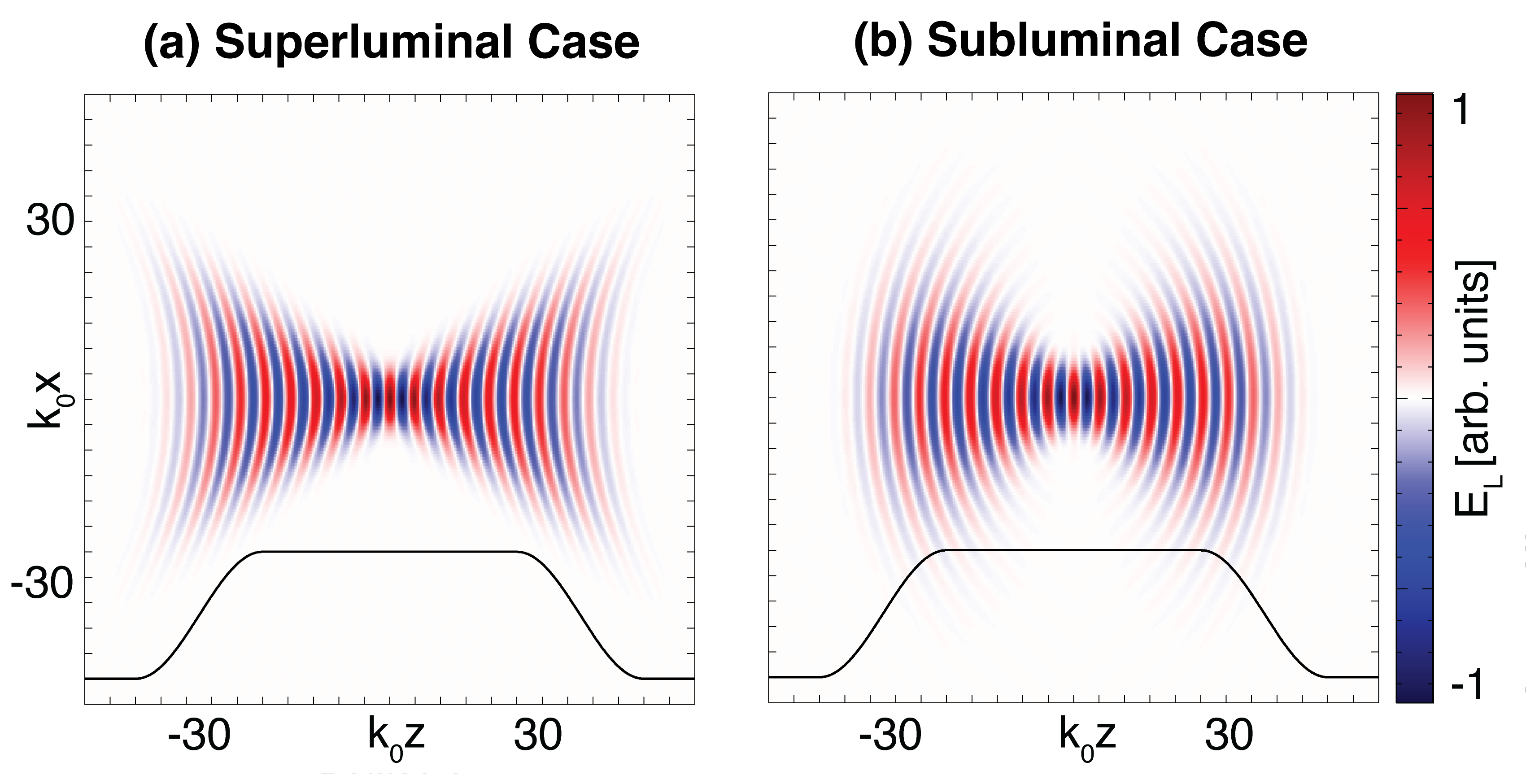}
\caption{Two examples of spatio-temporal pulses obtained with the discrete implementation in OSIRIS. In (a), we have an example of the shape of a superluminal pulse with $v_f= 1.5c$. In (b), we have an example of the shape of a subluminal pulse with $v_f= 0.5c$. The blue-red colour-scale represents the electric field, and the black line corresponds to the pulse's full longitudinal envelope $f_\parallel(z)$ in configuration space.} 
\label{fig:pulses_example}
\end{figure}

The longitudinal spectral function $f_\parallel(\kappa)$ is obtained by prescribing a desired temporal envelope and computing its Fourier transform with respect to the variable $\kappa$. In this work, we employ a flattened longitudinal profile (for example, a squared-sine envelope with a rapid rise and a long flat-top region) in order to produce a nearly constant on-axis intensity over an extended propagation distance. Figure~\ref{fig:pulses_example} shows representative examples of superluminal and subluminal space–time pulses constructed using this procedure, together with the temporal envelope employed.
\subsection{Discrete implementation in OSIRIS}
\label{subsec:numerical_init}

We now briefly describe the OSIRIS implementation of the method introduced in section~\ref{subsec:st_pulses}, originally developed by R.~Almeida~\cite{Almeida2023NonParaxialPIC}. The algorithm constructs the electromagnetic fields by evaluating the superposition in equation~\eqref{eq:plane_wave_superposition}, with the continuous spectral weight $f_k$ mapped onto a discrete grid in $k$-space. The fields are obtained by summing the resulting set of exact vacuum modes (for example, plane waves in Cartesian geometry or Bessel modes in cylindrical geometry) over the entire spatial domain at $t=0$.

The discretised implementation of the spectral weight introduces two main challenges. The first is that the initialisation is performed in Fourier space and, when transformed into configuration space, the simulation domain must be sufficiently large to contain the entire pulse. One possible way to address this issue is to continuously inject waves from the simulation boundaries, as will be discussed in detail in section~\ref{subsec:wallinjectionadvantage}.

The second challenge is that discretising $k$-space naturally introduces periodicity in configuration space, which may manifest in the simulation domain if not properly controlled. This issue is addressed in the following section.

\subsection{Accuracy constraints of the spectral initialisation}
\label{subsec:periodicity}

As discussed in section~\ref{subsec:numerical_init}, the numerical implementation relies on sampling the continuous spectral weight $f_k(\mathbf{k})$ on a finite and discretised set of modes with spacing $\Delta k_i$, where $i=(x,y,z)$. As in any discrete Fourier-type representation, this sampling leads to a periodic reconstruction of the electromagnetic fields in configuration space, with repetition lengths given by
\begin{equation}
T_{\mathrm{rep},i} = \frac{2\pi}{\Delta k_i}.
\end{equation}
These periodic replicas are numerical artefacts and may overlap with the physical pulse inside the simulation domain if not properly controlled.

To avoid unphysical effects in the longitudinal direction, the repetition length must exceed the size of the simulation window. This requires
\begin{equation}
T_{\mathrm{rep},z} > L_{\mathrm{box},z},
\end{equation}
so that no longitudinal replica enters the computational domain during the time interval of interest.

In the transverse directions, the repetition length must be sufficiently large to prevent overlap between the physical pulse and its periodic replicas throughout propagation. The quantity $T_{\mathrm{rep},(x,y)}$ represents the distance in configuration space between adjacent replicas of the pulse introduced by the discrete sampling in $k$-space. To avoid unphysical interference, this distance must exceed the maximum transverse extent of the pulse within the simulation domain.

For subluminal focal velocities $v_f<c$, the transverse evolution of the pulse can be estimated using the paraxial Gaussian-beam model~\cite{PhysRevA.107.013513, Almeida2023NonParaxialPIC},
\begin{equation}
w(\xi) = w_0 \sqrt{1 + \left( \frac{2\xi}{k_0 w_0^2 (1 - v_f/c)} \right)^2},
\label{eq:waist}
\end{equation}
where $\xi = z - v_f t$ is the co-moving coordinate.  

The transverse evolution of the pulse is described by the waist function $w(\xi)$, where $\xi = z - v_f t$ is the co-moving coordinate. At the focus, $w(0)=w_0$, while away from the focus, the pulse diffracts and its transverse size increases according to equation~\eqref{eq:waist}. Since the simulation domain is finite, the pulse only occupies a limited longitudinal interval $\xi$, determined by the boundaries of the computational box and the location of the focal point. The largest transverse size attained by the pulse inside the simulation window is therefore
\begin{equation}
w_{\max} = w(\xi_{\max}),
\end{equation}
which corresponds to the maximum beam radius that appears anywhere in the simulation.

Focusing on the transverse repetition length $T_{\mathrm{rep},(x,y)}$, artificial replicas of the pulse appear at transverse positions $x = \pm T_{\mathrm{rep},x}$ and $y = \pm T_{\mathrm{rep},y}$. To avoid unphysical overlap between the physical pulse and its replicas, the distance between neighbouring replicas must exceed the transverse size of the pulse throughout its evolution. Since the pulse occupies approximately the region $|x| \lesssim w_{\max}$, a sufficient geometric condition to prevent overlap is
\begin{equation}
T_{\mathrm{rep},(x,y)} \gtrsim 2 w_{\max}.
\end{equation}
This criterion ensures that, even at its widest point, the pulse remains spatially isolated from its periodic replicas. If this condition is violated, the replicas overlap with the main pulse, leading to artificial interference patterns and transverse modulations of the field.

\begin{figure}[t]
 \centering
 \includegraphics[width=0.9\textwidth]{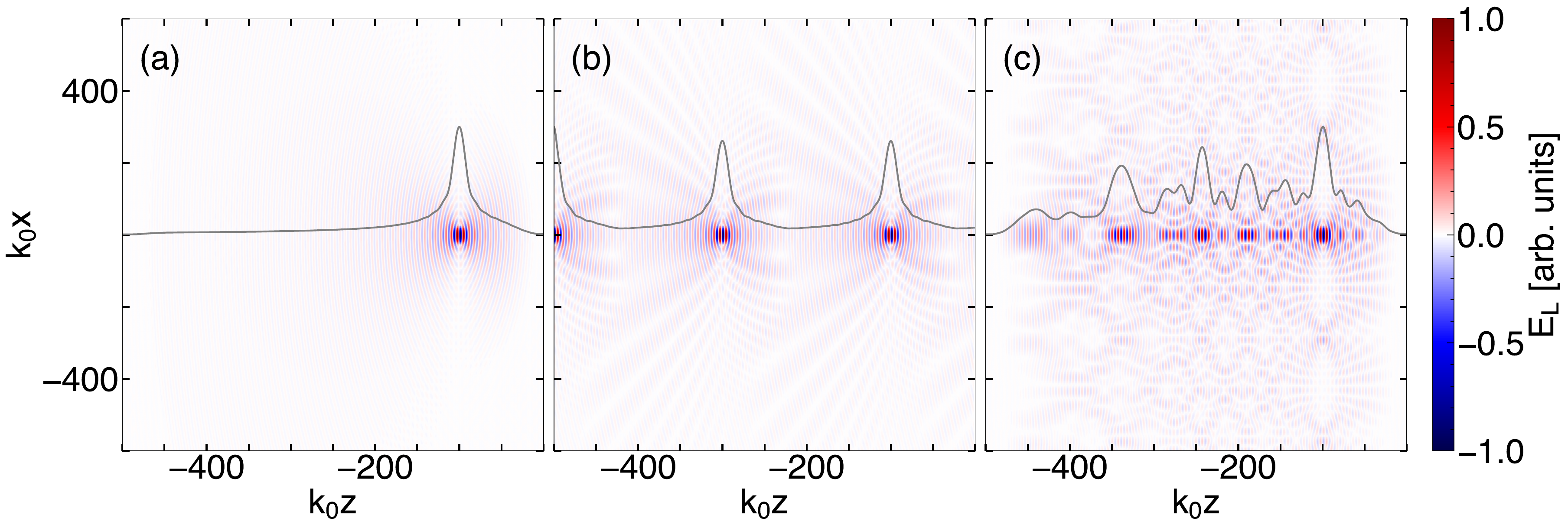}
 \caption{Comparison of the injected pulse electric field for three choices of the longitudinal and transverse repetition lengths used in the spectral reconstruction. (a) $k_0T_{\mathrm{rep},z}=700$ and $k_0T_{\mathrm{rep},(x,y)}=1000$, for which the reconstructed pulse is smooth and free of aliasing artefacts. (b) $k_0T_{\mathrm{rep},z}=200$ and $k_0T_{\mathrm{rep},(x,y)}=100$, where the longitudinal repetition length is too small, producing spurious replicas along the propagation axis. (c) $k_0T_{\mathrm{rep},z}=700$ and $k_0T_{\mathrm{rep},(x,y)}=200$, where the transverse repetition length is insufficient, and replicas enter the simulation domain, leading to distorted, unphysical structures.}
 \label{fig:periodicity}
\end{figure}
The impact of these constraints is illustrated in figure~\ref{fig:periodicity} for a laser pulse with parameters $\omega_0=10\,\omega_p$, $a_0 \equiv e E_0/\omega_0 m_e c =3$, $w_0=c/\omega_p$, and $v_f=0.96c$, where $E_0$ is the laser electric field maximum amplitude, $e$ is the elementary charge, and $m_e$ is the electron mass. Figure~\ref{fig:periodicity}(a) shows an optimised configuration in which both longitudinal and transverse repetition lengths satisfy the above criteria, resulting in a smooth pulse free from replica overlap. In contrast, figure~\ref{fig:periodicity}(b) shows the onset of artificial longitudinal modulations when $T_{\mathrm{rep},z}$ is too small, while figure~\ref{fig:periodicity}(c) illustrates the analogous transverse distortion that arises when $T_{\mathrm{rep},(x,y)}$ is insufficient compared with the pulse waist.

In summary, an artefact-free implementation can be obtained by first computing the maximum transverse size $w_{\max}$ of the pulse inside the simulation window using equation~\eqref{eq:waist}. The transverse repetition lengths should then be chosen such that $T_{\mathrm{rep},(x,y)} \ge 2 w_{\max}$, while the longitudinal repetition length must satisfy $T_{\mathrm{rep},z} > L_{\mathrm{box}}$.

\section{Simulation Design and Physical Validation}
\label{sec:design_validation}
The spectral geometry of the spatio-temporal pulses introduced in section~\ref{sec:theory_and_numerics} couples the driver’s transverse scale to the longitudinal extent of its high-intensity region. This differs from conventional Gaussian drivers, for which the spot size $w_0$ and pulse duration $\tau$ are usually treated as independent design parameters. As a result, resonant wakefield excitation does not follow the standard scaling laws directly and instead requires a dedicated exploration of the driver parameter space $(\omega_0, w_0, v_f, a_0)$. 

Moreover, because the prescribed focal velocity $v_f$ is decoupled from the envelope group velocity, long-distance propagation introduces practical numerical constraints. The relative drift between the moving focus and the finite envelope sets a survival length, while the broad spectral support required for $v_f\neq c$ leads to substantial transverse spreading. Together, these effects can make full-domain initialisation increasingly expensive in the subluminal regime.

This section addresses both the physical and numerical requirements. Section~\ref{subsec:pulseshaping} analyses how $v_f$ and the driver geometry constrain the resonant conditions for wake excitation and derives practical scaling guidelines for parameter selection. Section~\ref{subsec:wallinjectionadvantage} then introduces and benchmarks continuous wall injection, showing how it alleviates the domain-size constraints and establishing the criteria needed for efficient and stable particle-in-cell modelling of these spatially extended drivers.

\subsection{Optimisation of spatio-temporal drivers for wakefield excitation}
\label{subsec:pulseshaping}

Efficient wakefield generation relies on matching the laser-pulse dimensions to the characteristic plasma scales. Transversely, the energy should be concentrated within a region comparable to the plasma skin depth, $k_p^{-1}=c/\omega_p$, where $\omega_p=(n_0e^2/\epsilon_0 m_e)^{1/2}$ is the electron plasma frequency, $n_0$ is the plasma density, and $\epsilon_0$ is the vacuum permittivity. Longitudinally, resonant excitation typically requires the driver length to be of the order of half the wake wavelength~\cite{malka2020cas}. Two points are important in the present context. First, for the spatio-temporal pulses considered in this work, the transverse and longitudinal profiles are intrinsically coupled and cannot be tuned independently. Second, when the wake is analysed in the co-moving coordinate $\xi=z-v_f t$, the apparent wake period in $\xi$ depends on the focal velocity $v_f$. In particular, a plasma oscillation at $\omega_p$ corresponds to the spatial period
\begin{equation}
\lambda_{\mathrm{wake}}(v_f)=\frac{2\pi v_f}{\omega_p},
\end{equation}
so the longitudinal matching condition in $\xi$ is $L_{\text{pulse}}\sim \lambda_{\mathrm{wake}}/2$.

The corresponding field envelope, valid in the paraxial limit ($k_0 w_0 \gg 1$), can be written as
\begin{equation}
E(r,\xi) \propto \frac{w_0}{w(\xi)} \exp\!\left(-\frac{r^2}{w(\xi)^2}\right),
\label{eq:field}
\end{equation}
where $w(\xi)$ is given by equation~\eqref{eq:waist}. In the subluminal regime ($|v_f|<c$), equation~\eqref{eq:waist} has the same functional form as the diffraction of a vacuum Gaussian beam, but with the propagation coordinate replaced by the co-moving variable $\xi$. This motivates defining an effective Rayleigh length in $\xi$,
\begin{equation}
\xi_R \equiv \frac{k_0 w_0^2}{2}\left(1-\frac{v_f}{c}\right),
\end{equation}
so that $w(\xi)=w_0\sqrt{1+(\xi/\xi_R)^2}$. Within this paraxial model, the on-axis intensity scales as
\begin{equation}
I(0,\xi)\propto \left[1+\left(\frac{\xi}{\xi_R}\right)^2\right]^{-1},
\end{equation}
which is Lorentzian in $\xi$ with full width at half maximum
\begin{equation}
L_{\mathrm{FWHM}} = 2\xi_R = k_0 w_0^2\left(1-\frac{v_f}{c}\right).
\end{equation}

For a standard Gaussian beam, the Rayleigh length characterises the spatial decay of the peak intensity with propagation distance. Here, by contrast, $\xi_R$ sets the extent of the high-intensity region in the co-moving coordinate. Since a Lorentzian profile has long tails, we characterise the effective driver length by the $1/\mathrm{e}^2$ intensity width ($\mathrm{e}$ is the Euler number),
\begin{equation}
L_{\text{pulse}} \equiv L_{1/\mathrm{e}^2} 
= \sqrt{\mathrm{e}^2-1}\,L_{\mathrm{FWHM}}.
\end{equation}
Imposing the longitudinal matching condition $L_{\text{pulse}}\sim \lambda_{\mathrm{wake}}/2$ then yields the estimate
\begin{equation}
k_0 w_0 \approx
\left[
\frac{\pi}{\sqrt{\mathrm{e}^2-1}}
\frac{\omega_0}{\omega_p}
\frac{v_f/c}{1-v_f/c}
\right]^{1/2}.
\label{eq:k0w0_optimal}
\end{equation}

\begin{figure}[tb]
    \centering
    \includegraphics[width=0.8\textwidth]{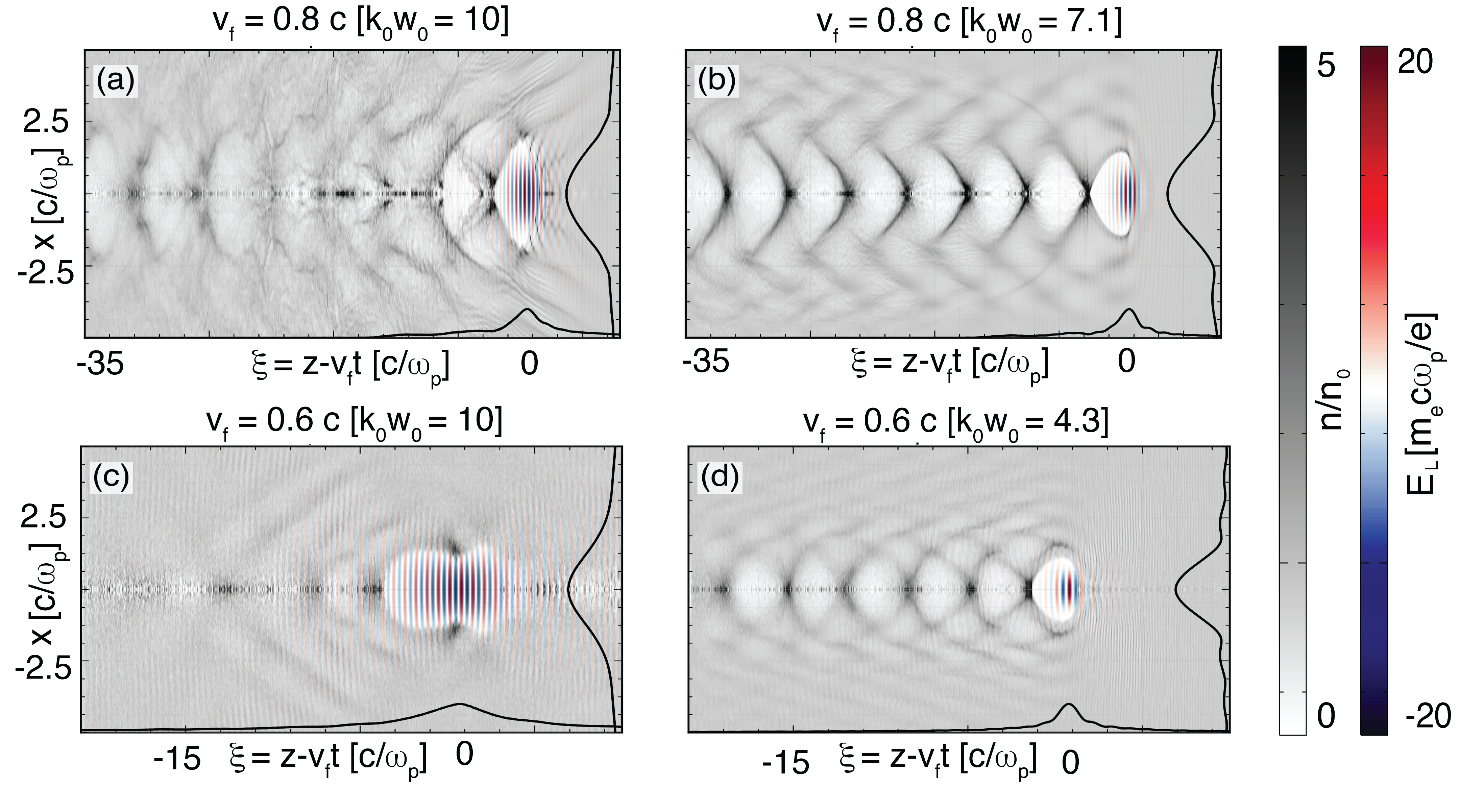}
    \caption{Wakefield excitation for different choices of $v_f$ and $w_0$. Panels (a) and (b) show the case $v_f=0.8c$: for (a) $k_0w_0=10$ the driver is too long and overlaps with most of the first plasma period, whereas for (b) $k_0w_0=7.1$  the driver length is better matched and a clearer wake structure is obtained. The same trend is more pronounced in panels (c) and (d) for $v_f=0.6c$, comparing (c) $k_0w_0=10$  with the tighter-focus case (d) $k_0w_0=4.3$. The greyscale shows the electron density, and the blue--red colour scale shows the laser electric field.  All cases use $\omega_0/\omega_p=10$ and $a_0=2.2$.}
    \label{fig:velocity_scan}
\end{figure}
This longitudinal estimate should be complemented by the usual transverse matching condition for wakefield excitation. At the pulse centre ($\xi=0$), the transverse profile reduces to a Gaussian with spot size $w_0$, so efficient coupling in conventional wakefield theory typically requires $k_p w_0 \sim \mathcal{O}(1)$. Since the longitudinal resonance condition and the transverse matching condition both constrain the same parameter $w_0$, it is not, in general, possible to satisfy both simultaneously. Efficient operation therefore requires working in a regime where the two estimates are compatible, or prioritising one constraint and accepting a compromise in the other.

Figure~\ref{fig:velocity_scan} illustrates how the interplay between $w_0$ and $v_f$ can produce either weakly resonant or near-optimal wakefields for fixed parameters $\omega_0/\omega_p = 10$ and $a_0 = 2.2$. Figures~\ref{fig:velocity_scan}(a) and~(b) correspond to $v_f = 0.8c$. In panel (a), the choice $k_0 w_0 = 10$ is not optimal, whereas in panel (b) we use the value predicted by equation~\eqref{eq:k0w0_optimal}, $k_0 w_0 \approx 7.1$. The latter case exhibits a markedly cleaner plasma response (shown in grey) to the ST driver. Similarly, figures~\ref{fig:velocity_scan}(c) and~(d) correspond to $v_f = 0.6c$. For $k_0 w_0 = 10$ [panel (c)], the driver is noticeably longer than the plasma response, while using the predicted optimal value $k_0 w_0 \approx 4.3$ [panel (d)] restores a well-defined wake structure even at this lower focal velocity.
\begin{figure}[b]
\centering \includegraphics[width=0.7\textwidth]{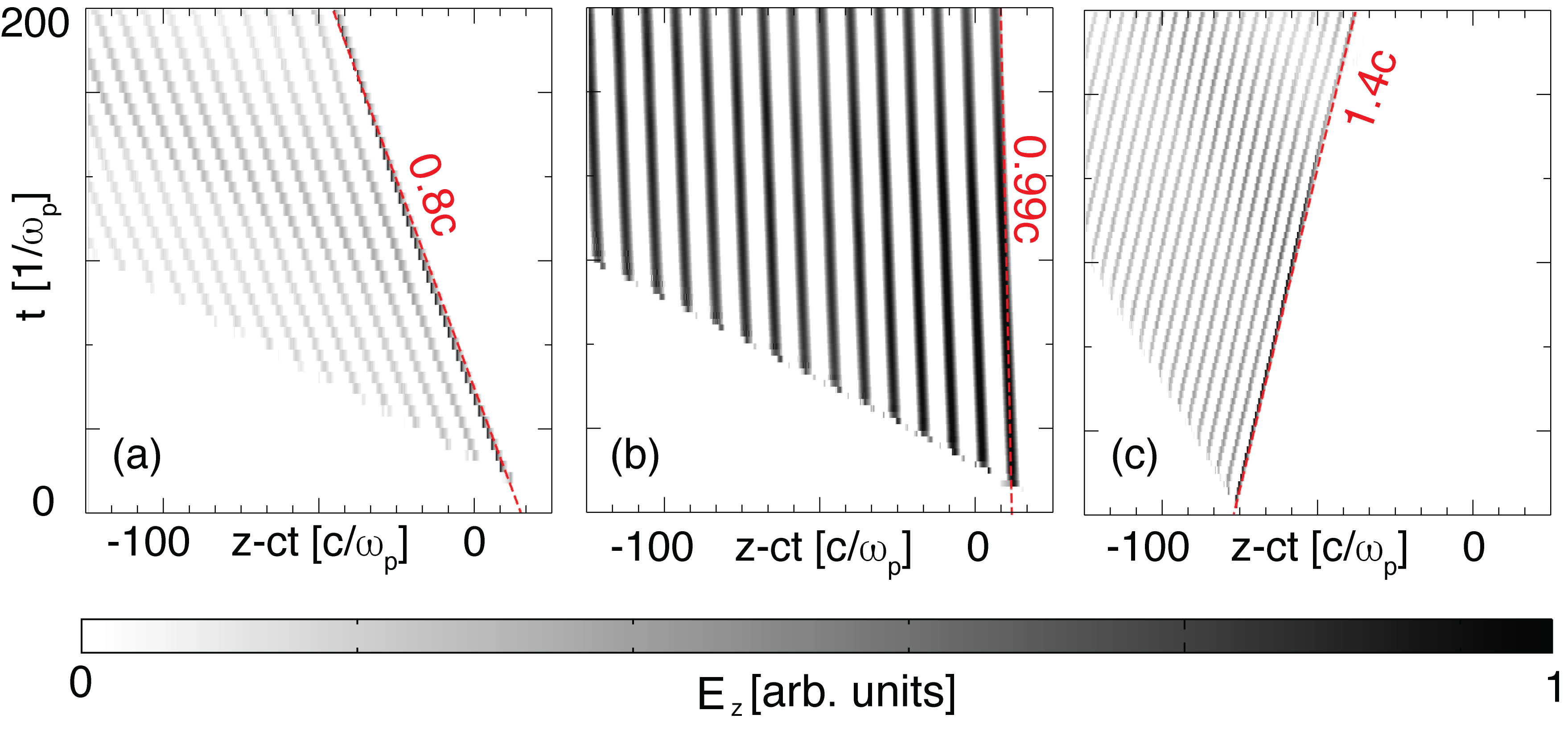}
\caption{Waterfall plots of the accelerating electric-field amplitude for different focal velocities: (a) $v_f=0.8c$, (b) $v_f=0.99c$, and (c) $v_f=1.4c$.}
\label{fig:waterfall_differentvf}
\end{figure}

However, this optimisation strategy may introduce trade-offs. Experimentally, generating tightly focused broadband wave packets requires large-aperture optics. In addition, a smaller spot size produces a narrower transverse wake structure, which reduces the acceptance for a witness beam. Despite these constraints, the wakefield remains stable over extended propagation distances when the appropriate spatio-temporal coupling is employed, as illustrated in figure~\ref{fig:waterfall_differentvf}, which shows waterfall plots of the longitudinal electric field for spatio-temporal drivers with different focal velocities, together with the corresponding wake evolution that closely follows the imposed focal velocity.

\begin{figure}[htb]
    \centering
    \includegraphics[width=1.0\textwidth]{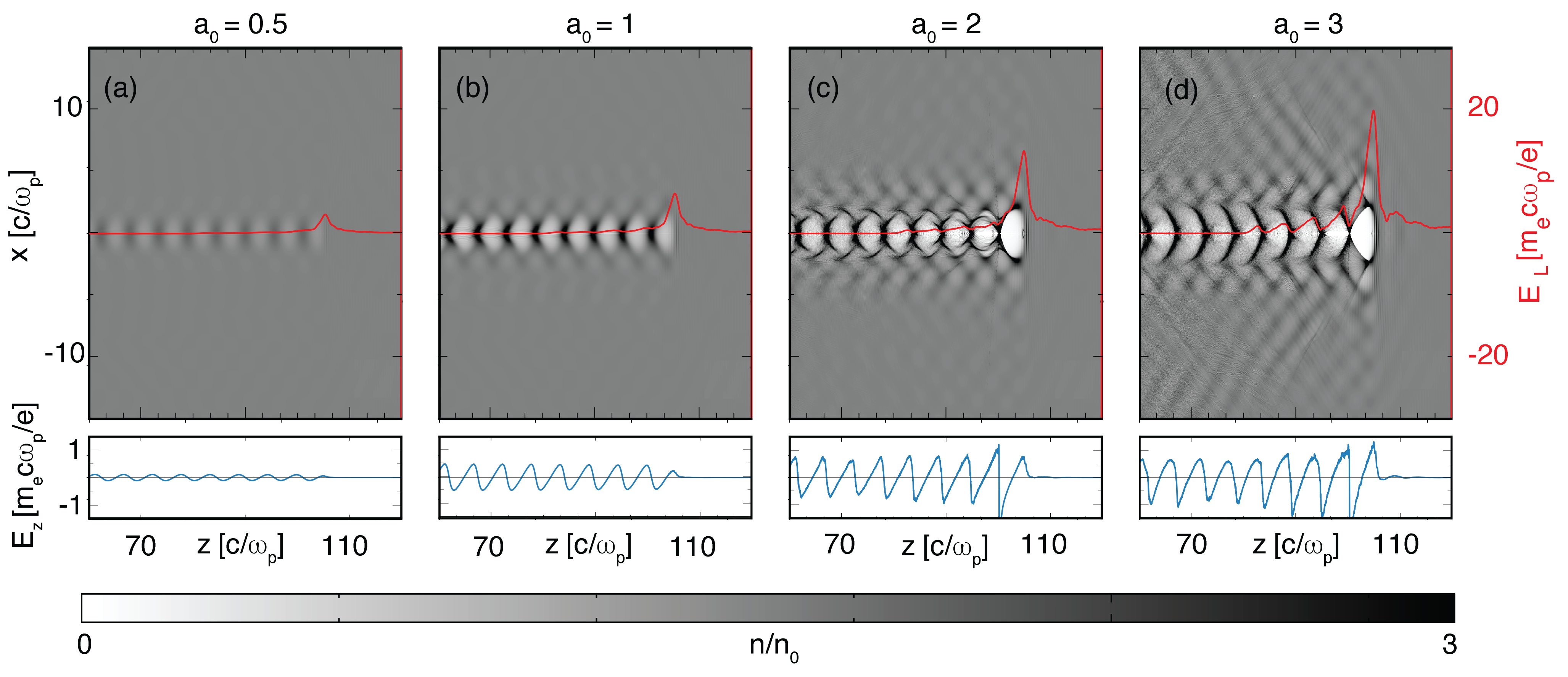}
    \caption{Wakefield excitation for a fixed driver with $v_f=0.96c$ and $k_0w_0=17$, for different values of $a_0$. The upper panels show the electron density response and the on-axis laser electric-field envelope, while the lower panels show the on-axis accelerating field. (a) $a_0=0.5$, (b) $a_0=1$, (c) $a_0=2$, and (d) $a_0=3$.}
    \label{fig:a0_scan}
\end{figure}
Once the spatio-temporal geometry is optimised for resonance, the wakefield amplitude can be further increased by raising the driver strength $a_0$. As illustrated in figure~\ref{fig:a0_scan}, increasing from (a) $a_0=0.5$, (b) $a_0=1$, (c) $a_0=2$, and (d) $a_0=3$ leads to progressively larger accelerating fields and to changes in the wake structure, going from the linear to the non-linear regime. The wakefield patterns obtained, particularly in panel~(d), are qualitatively similar to those reported in recent experiments demonstrating wakefield generation driven by structured light~\cite{Liberman2025}.

In summary, the paraxial model provides a practical framework to select spatio-temporal pulse parameters that yield near-resonant and stable wakefield excitation. Because the longitudinal and transverse properties of the driver are intrinsically coupled, resonance cannot be enforced by tuning the pulse duration alone. In the subluminal regime, reducing the focal velocity $v_f$ increases the effective longitudinal extent of the high-intensity region in the co-moving coordinate, which can violate the resonance condition. Maintaining resonant excitation, therefore, requires compensating for this elongation by reducing the spot size $w_0$ according to the scaling in equation \eqref{eq:k0w0_optimal}. This provides a simple design strategy to identify suitable $(w_0,v_f)$ combinations before performing full numerical optimisation.

\subsection{Simulation domain size and wall injection for spatio-temporal pulses}
\label{subsec:wallinjectionadvantage}

In conventional laser–wakefield simulations, the driver is typically a short pulse whose longitudinal extent is small compared with its diffraction length. A moving window that follows the driver at (approximately) the speed of light can therefore contain both the laser and the plasma response within a compact computational domain.

\begin{figure}[t]
    \centering
    \includegraphics[width=0.7\textwidth]{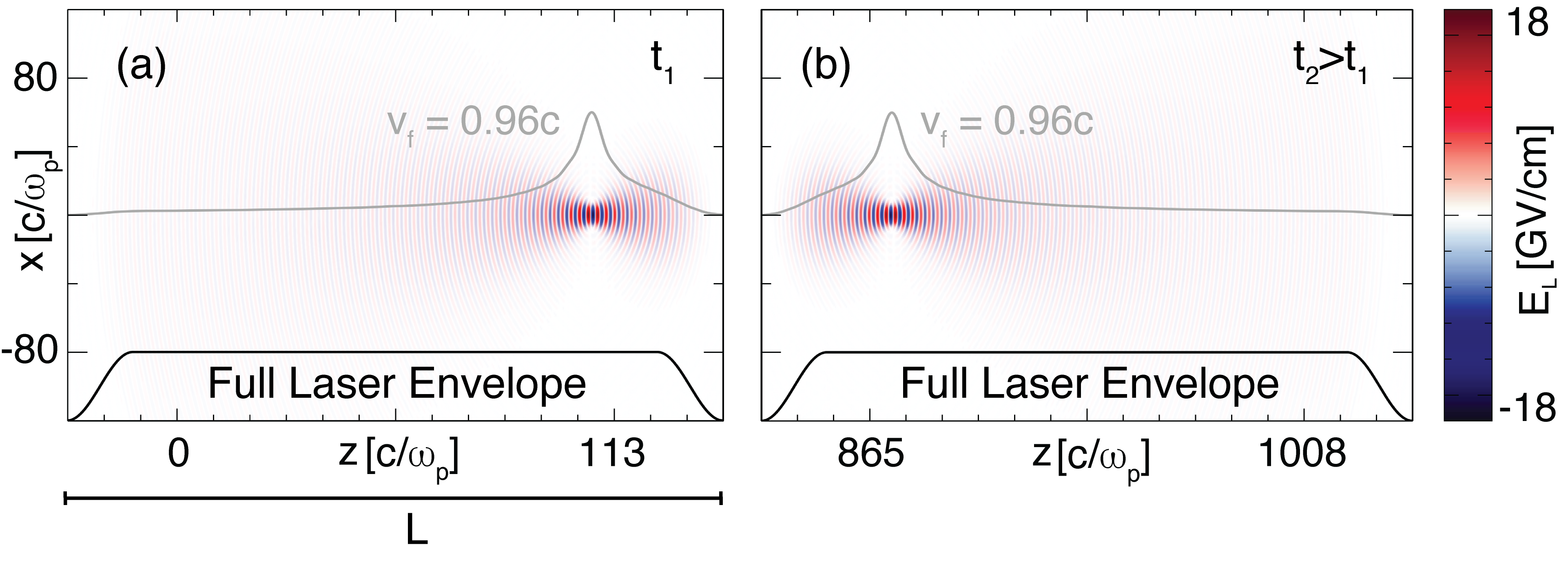}
    \caption{Schematic of a subluminal ST pulse showing the full laser envelope (black) and the localised high-intensity region (grey line; blue--red shading). Because the envelope propagates at $c$ while the focus advances at $v_f<c$, the intensity peak slips backwards within the envelope between (a) $t_1$ and (b) $t_2>t_1$. Once the peak reaches the trailing edge of the envelope, the driver can no longer sustain an efficient wake.}
    \label{fig:envelopeslipping}
\end{figure}

Velocity-controlled ST pulses differ in two key ways, with direct consequences for the required box size. First, the driver is not a short wave packet: the laser energy is distributed over a comparatively long envelope, while the high-intensity peak is formed by a constructive-interference condition that propagates through this envelope at a prescribed focal velocity $v_f$. In a standard moving window that advances at $c$ (and thus tracks the envelope), the envelope remains approximately stationary, whereas the intensity peak drifts through it at the relative speed $c-v_f$ in subluminal cases. This slippage is illustrated in figure~\ref{fig:envelopeslipping}, where the focus slips from the leading side of the envelope towards the trailing side as propagation proceeds.

This relative motion implies that the high-intensity peak has a finite lifespan within a given envelope. We quantify this effect by defining a survival length $L_{\mathrm{surv}}$, namely the maximum distance over which the focus remains contained inside an envelope of longitudinal extent $L$. For a subluminal pulse with the focus initially near the leading edge, the survival time is $\Delta t_{\mathrm{surv}} \simeq L/(c-v_f)$, which gives
\begin{equation}
L_{\mathrm{surv}} \simeq \frac{v_f}{c-v_f}\,L.
\label{eq:survival_length}
\end{equation}
In wakefield-driven acceleration, $L_{\mathrm{surv}}$ provides an upper bound on the propagation distance over which the driver can maintain a stable wake structure purely from this geometric constraint. In practice, achieving long interaction lengths at focal velocities far from $c$ therefore requires either a correspondingly long envelope (and thus a larger longitudinal domain).

 Constructing a pulse with a prescribed focal velocity $v_f\neq c$ requires a broad spectral support with significant transverse wavevector content. In the subluminal regime, this manifests as a large effective divergence of the envelope, so that the pulse occupies a wider transverse extent than a conventional Gaussian driver of comparable $w_0$. Consequently, avoiding artificial truncation generally demands a larger transverse domain.

These considerations must be reconciled with the numerical constraints discussed in section~\ref{subsec:periodicity}. The longitudinal repetition length introduced by spectral discretisation may need to be increased as the envelope length and interaction distance grow.

The simulation domain requirements depend on the chosen ST pulse parameters and can be estimated from simple geometric considerations. For subluminal pulses with a Gaussian transverse profile, the envelope expands transversely as it propagates. For a pulse with spot size $w_0$ and focal velocity $v_f$, an effective divergence half-angle can be estimated as
\begin{equation}
\tan\,\theta_{\mathrm{div}} \approx \frac{2}{k_0 w_0 \left(1 - v_f/c\right)}.
\label{eq:thetadiv}
\end{equation}

\begin{figure}[t]
    \centering
    \includegraphics[width=0.7\textwidth]{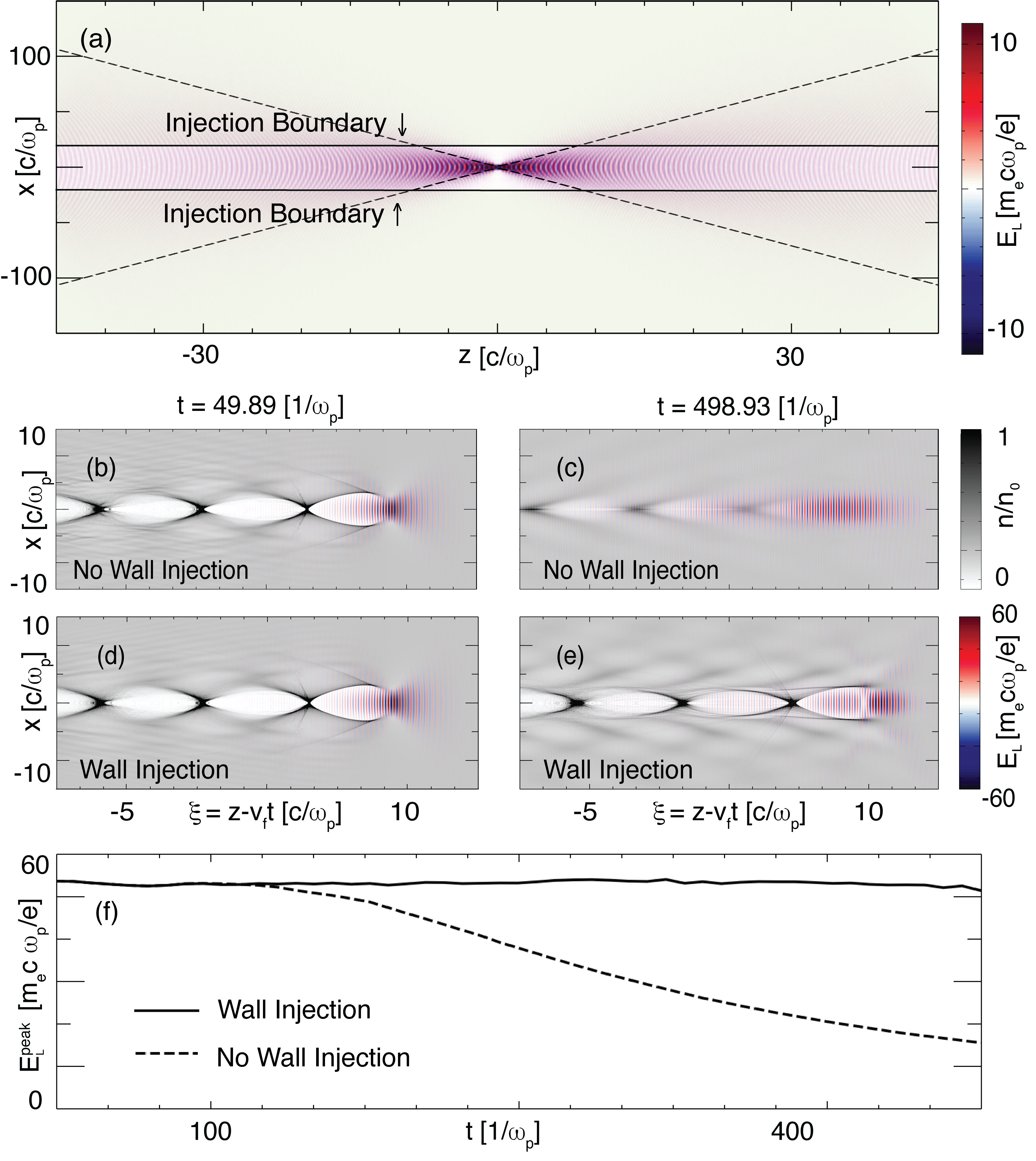}
    \caption{Reduced-domain simulations of a subluminal ST driver with and without wall injection. (a) schematic of the injected field and boundary locations. (b,d) early-time snapshots in the co-moving coordinate $\xi$ without and with wall injection, respectively. (c,e) late-time snapshots showing wake degradation without injection and sustained excitation with injection. (f) peak intensity in vacuum versus time, confirming that wall injection preserves the intended propagation-invariant behaviour in the reduced domain. Greyscale shows the electron density and the blue--red scale shows the laser electric field.
}
    \label{fig:sstpulse_wallinjection}
\end{figure}

Figure~\ref{fig:sstpulse_wallinjection} shows an example with $v_f = 0.97c$, $k_0 w_0 = 14$, and $\omega_0/\omega_p = 10$. Using equation~\eqref{eq:thetadiv}, this corresponds to $\theta_{\mathrm{div}} \approx 67^\circ$, indicated by the dashed black lines in figure~\ref{fig:sstpulse_wallinjection}(a). For a Gaussian transverse profile, this aperture contains approximately 86\% of the pulse energy~\cite{saleh1991chapter3}. The transverse domain size must therefore be chosen to accommodate the resulting expansion over the propagation distance $z_{\mathrm{prop}}$, thereby avoiding truncation or boundary interactions of the driver.

In a quasi-3D geometry, where the transverse coordinate ranges from $r\in[0,L_\perp]$, a practical guideline is
\begin{equation}
L_\perp \gtrsim 4w_0 + z_{\mathrm{prop}}\tan\theta_{\mathrm{div}},
\label{eq:Lperp_guideline}
\end{equation}
so that the relevant initial envelope extent (taken here as $4w_0$) and its subsequent expansion remain contained within the computational domain. For a propagation throughout the full survival time, a conservative choice is $z_{\mathrm{prop}} \sim c\,\Delta t_{\mathrm{surv}}$, which typically implies a large transverse domain.

Fortunately, the transverse-domain requirements can be relaxed by using a boundary-injection approach. Instead of initialising the full field distribution throughout the simulation domain at $t=0$, OSIRIS can inject the driver continuously through the transverse boundaries by evaluating the superposition in equation~\eqref{eq:plane_wave_superposition} at each time step. In this scheme, the angular spectrum required to form the ST pulse is supplied dynamically from the walls, so the simulation does not need to contain the full laterally expanding envelope at all times. This allows the transverse domain to be reduced substantially while preserving the correct field structure in the interaction region and maintaining consistency with the fields obtained from full-domain initialisation.

Figure~\ref{fig:sstpulse_wallinjection}(a) illustrates that the transverse domain can be reduced substantially by focusing the computational region on the portion of the pulse that is most relevant for the laser–plasma interaction, i.e.\ where the intensity is sufficiently high. Figures ~\ref{fig:sstpulse_wallinjection}~(b) and~(c) show simulations performed in this smaller box without boundary injection. While the early-time response is captured, the wakefield at later times [panel~(c)] becomes markedly weaker because the pulse energy is no longer represented correctly: the laterally diverging components required to sustain the moving focus are truncated by the boundaries.

Figures~\ref{fig:sstpulse_wallinjection}(d) and~(e) show the corresponding simulations using wall injection. At early times, the results are indistinguishable from the case without wall injection (compare panels~(b) and~(d)). At later times, wall injection sustains a wakefield amplitude comparable to that obtained initially, although deviations from the early-time structure (panel~(d)) emerge due to laser--plasma interaction effects. In particular, the plasma modifies the field components that would otherwise converge from the transverse ``wings'' to form the interference pattern that defines the moving focus. This highlights an important point: although ST pulses can be designed to be shape invariant in vacuum, their structure can be altered by interaction with plasma.

To corroborate that the wall-injection algorithm preserves the intended propagation-invariant intensity profile, figure~\ref{fig:sstpulse_wallinjection}(f) shows the evolution of the peak laser intensity in vacuum, with and without wall injection. Without injection, the peak intensity decays as the pulse footprint expands beyond the reduced domain, whereas with wall injection it remains approximately constant. This demonstrates that, in vacuum, wall injection maintains the desired field structure while enabling substantially smaller transverse domains.

To quantify the practical benefit of wall injection in a realistic PIC setup, we benchmark the wall-injection algorithm against the conventional approach in which the full pulse is initialised throughout the domain at $t=0$. Because wall injection evaluates the analytic field at the transverse boundaries throughout the run, it introduces an approximately constant computational overhead per time step, increasing the time per iteration by a factor of $\sim 1.5$--2. This overhead is typically more than compensated by the reduction in domain size enabled by wall injection: without injection, containing the laterally expanding field can require a transverse domain that is $\sim 10$--100 times larger, depending on the pulse parameters and the interaction length. Since the overall computational workload scales with the number of degrees of freedom advanced in time, reducing the domain size directly reduces the total computational resources required. In quasi-3D, the cost scales approximately with the number of grid cells, while in full 3D it scales as $N_\perp^2$ in the transverse resolution. Consequently, wall injection can reduce both the memory footprint and the overall computational cost, often by one to two orders of magnitude relative to otherwise equivalent simulations.

As a practical guide, simulations of subluminal ST drivers should be sized using the survival time $\Delta t_{\mathrm{surv}}\simeq L/(c-v_f)$ and the effective divergence $\tan\theta_{\mathrm{div}}\approx 2/[k_0w_0(1-v_f/c)]$. In quasi-3D geometry, choose the radial extent to satisfy $L_\perp \gtrsim 4w_0+z_{\mathrm{prop}}\tan\theta_{\mathrm{div}}$, with $z_{\mathrm{prop}}$ set by the intended interaction length (a conservative choice is $z_{\mathrm{prop}}\sim c\,\Delta t_{\mathrm{surv}}$ when propagating over the full survival time). If these requirements make the transverse domain impractical, enable wall injection, which supplies the angular spectrum at the boundaries and allows the interaction region to be simulated in a much smaller box.
\section{Conclusions}
\label{sec:conclusions}

This work presented a practical workflow for modelling velocity-controlled spatio-temporal (ST) laser drivers in particle-in-cell simulations. We rely on a Maxwell-consistent spectral construction that defines ST pulses as a superposition of exact vacuum solutions, ensuring the correct spatio-temporal correlations. We then described how to map this continuous formulation onto a discrete $k$-space grid suitable for numerical initialisation.

A critical numerical consequence of discretising the spectrum is the appearance of real-space replicas, whose periodicity is determined by the $k$-space sampling. We explicitly address this issue by providing simple, geometry-based criteria for selecting the longitudinal and transverse repetition lengths. These criteria ensure that replica pulses do not enter the interaction region during the simulation, making the spectral method reliable in practical wakefield setups.

We have shown that the ST spectral geometry intrinsically couples the driver’s transverse scale and the longitudinal extent of its high-intensity region. As a result, wakefield optimisation cannot be performed by varying the pulse duration independently of the spot size, as is typically done for conventional Gaussian drivers. In the subluminal regime, decreasing $v_f$ elongates the effective high-intensity region in the co-moving coordinate; we have demonstrated that near-resonant excitation can be recovered by reducing $w_0$ according to the scaling in equation~\eqref{eq:k0w0_optimal}, and we have validated this design rule with PIC simulations across representative parameter scans.

We have addressed the numerical constraints that arise when $v_f\neq c$. First, the slip between the envelope (advancing at c) and the moving focus introduces a survival length, limiting how long a stable wake can be sustained within a finite envelope. Second, the broad angular content required for subluminal pulses leads to substantial transverse expansion. These geometric effects make full-domain initialisation increasingly expensive for long subluminal runs. We have therefore employed a continuous wall-injection scheme that supplies the required self-consistent fields dynamically through the transverse boundaries. We have shown that wall injection preserves the intended propagation-invariant behaviour in vacuum and sustains the driven wake in reduced transverse domains, while also highlighting that laser--plasma interaction can modify the pulse structure relative to the vacuum design. Although wall injection introduces a modest per-step overhead, the reduction in domain size typically yields a net decrease in overall computational cost, often by one to two orders of magnitude for otherwise equivalent simulations.

Collectively, these results provide a concise recipe for the robust PIC modelling of velocity controlled spatio-temporal drivers drivers: (i) specify the driver through a Maxwell-consistent spectral decomposition; (ii) choose the $k$-space discretisation so that the associated repetition lengths exceed the relevant longitudinal and transverse window; (iii) select $(w_0,v_f)$ using the resonance scaling in equation~\eqref{eq:k0w0_optimal} to obtain strong and stable wake excitation; and (iv) for long subluminal propagation, use wall injection to avoid representing the full laterally expanding envelope inside the computational domain.
\appendix
\section{Appendix: Simulation setup}
\label{app:simulationsetup}
All simulations are performed with the OSIRIS azimuthal-mode decomposition algorithm in $(r,z$--$\varphi)$ geometry (quasi-3D)~\cite{lifschitz2009particle,davidson2015implementation}. In this approach, the computational grid is two-dimensional in $(r,z)$, while the dependence on the azimuthal angle $\varphi$ is represented analytically by expanding the fields and currents into a finite number of Fourier modes. In other words, there is no mesh in $\varphi$: non-axisymmetric structure is captured through the retained azimuthal harmonics, and the accuracy is controlled by the number of modes included. This provides a practical compromise between fully three-dimensional modelling and the lower cost of axisymmetric simulations.

We retain the cylindrically symmetric mode $m=0$ and the first non-axisymmetric mode $m=1$, which is sufficient to capture the laser polarisation and the main wake structure at reduced computational cost. The plasma is initialised with a uniform density $n_0$, preceded by a short vacuum-to-plasma density ramp of length $L_{\mathrm{ramp}}=4\,c/\omega_p$. Plasma electrons are sampled with 32 particles per cell. A moving window propagating at the speed of light is employed in all simulations presented in this manuscript.

For the parameter scans shown in figures~\ref{fig:waterfall_differentvf} and~\ref{fig:a0_scan}, the simulation domain is $L_z\times L_r = 100\,c/\omega_p \times 20\,c/\omega_p$ with $N_z\times N_r = 10000\times 400$, corresponding to $\Delta z = 0.01\,c/\omega_p$ and $\Delta r = 0.05\,c/\omega_p$. Absorbing boundary conditions are applied at the transverse edge of the domain using perfectly matched layers.

The laser driver is polarised along $E_y$. Its longitudinal envelope is a flat-top profile with squared-sine ramps on both sides (rise/fall length $5\,c/\omega_p$ and flat-top length $90\,c/\omega_p$).

\ack{This work was supported by FCT I.P. under Project 2024.07895.CPCA.A3 – DOI: https://doi.org/10.54499/2024.06987.CPCA.A3 – at MareNostrum 5 supercomputer, jointly funded by EuroHPC JU, Portugal, Turkey and Spain. This work was supported by the HE EuPRAXIA-PP under grant agreement No. 101079773. CB acknowledges the support of the Portuguese Science Foundation (FCT) Grant No. PRT/BD/152270/2021 (DOI: 10.54499/PRT/BD/152270/2021), RA FCT Grant No. UI/BD/154677/2022 (DOI: 10.54499/UI/BD/154677/2022), and TS FCT IPFN-CEEC-INST-LA3/IST-ID.}

\data{The data cannot be made publicly available upon publication because they are not available in a format that is sufficiently accessible or reusable by other researchers. The data that support the findings of this study are available upon reasonable request from the authors.}

\bibliographystyle{ieeetr} 
\bibliography{bibliography} 
\end{document}